\documentclass[letterpaper, 11pt]{article}

\usepackage{orcidlink}
\usepackage{hyperref}
\usepackage{mathtools}
\usepackage{array}
\usepackage{enumitem}
\usepackage{amsmath,amsfonts,amssymb,bm}
\usepackage{authblk}
\usepackage{titlesec}
\usepackage{pgfgantt,rotating}
\usepackage{float}
\usepackage{graphicx}
\usepackage{sidecap}
\usepackage{wrapfig}
\usepackage{bold-extra}
\usepackage{authblk}
\usepackage{appendix}
\usepackage{makecell}
\usepackage{dirtytalk}
\usepackage{placeins}
\usepackage[export]{adjustbox}
\usepackage{cite}
\usepackage{graphicx}
\usepackage{stackengine}
\setstackgap{L}{0pt}
\usepackage[margin=1in]{geometry}
\usepackage{graphicx}
\usepackage{subcaption}

\usepackage[percent]{overpic}

\newcommand{\beq}{\begin{equation}}
\newcommand{\eeq}{\end{equation}}
\newcommand{\bgqar}{\begin{eqnarray}}
\newcommand{\enqar}{\end{eqnarray}}
\newcommand{\bgqarn}{\begin{eqnarray*}}
\newcommand{\enqarn}{\end{eqnarray*}}
\newcommand{\bgary}{\begin{array}}
\newcommand{\enary}{\end{array}}

\graphicspath{{Figures/}}

\title{Koopman Spectral Analysis of Lithium-Ion Battery  Dynamics: SOC as a Marginally Stable Observable}
\title{Koopman Spectral Analysis of Lithium-Ion Battery Dynamics: State of Charge as a Marginally Stable Observable}

\author[1]{Bakhtiar Nafis \orcidlink{0009-0005-3795-5155}}
\author[1]{Khalid Mahmud Labib \orcidlink{0009-0007-4724-7747}}
\author[1]{Saad Waheed \orcidlink{0009-0002-8700-6140}}
\author[1,$\ast$]{Shabbir Ahmed \orcidlink{0000-0001-8296-6025}}

\affil[1]{Dynamical Systems and Signals Lab (DSSL),
Department of Mechanical Engineering,
South Dakota State University, Brookings, SD}

\date{
\small $^\ast$Corresponding author: \href{mailto:shabbir.ahmed@sdstate.edu}{shabbir.ahmed@sdstate.edu}
}

\begin{document}

\maketitle

\begin{abstract}

Accurate state-of-charge (SOC) estimation in lithium-ion batteries remains a challenge due to its strongly nonlinear electrochemical dynamics, operating-condition-dependent parameters, and aging-induced model drift. Conventional equivalent circuit model (ECM) and Kalman-filter-based estimators require repeated parameter identification, while purely data-driven methods sacrifice physical interpretability. We propose a Koopman operator theoretic SOC estimation framework that leverages dynamic mode decomposition with control (DMDc) and Hankel time-delay embedding to identify a linear representation of battery dynamics from input-output measurements. Terminal voltage and current measurements from hybrid pulse power characterization (HPPC) tests are lifted into a high-dimensional observable space via Hankel embedding, enabling linear approximation of the underlying nonlinear dynamics. DMDc identifies the state-transition operator directly from data, and eigen-decomposition of this operator reveals the intrinsic Koopman spectral structure of the battery. The SOC dynamics emerge as the slowest marginally stable mode, with eigenvalue nearest to the unit circle, consistent with the integrator-type pole implied by charge conservation. The corresponding modal coordinate provides a physically grounded SOC-sensitive observable, extracted without any explicit circuit parameter identification. The proposed framework reconstructs terminal voltage with an RMSE of 0.0131 V and estimates SOC with an RMSE of 0.0043\%, outperforming both Coulomb counting and the extended Kalman filter.

% \linebreak
\textbf{Keywords:} Lithium-ion batteries, State of Charge estimation, Dynamic Mode Decomposition with control, Koopman operator, Hankel embedding, Battery Management System

\end{abstract}

%------------------------------------------------------------------------------------------------------------
%                              Important conventions and symbols -- Acronyms
%------------------------------------------------------------------------------------------------------------

%------------------------------------------------------------------------------------------------------------

\newpage\pagebreak 

% \tableofcontents 

%------------------------------------------------------------------------------------------------------------
% Introduction
%------------------------------------------------------------------------------------------------------------

\section{Introduction} \label{sec:intro}

Lithium-ion batteries have become the dominant energy storage technology across a broad range of modern applications owing to their high energy density, long cycle life, low self-discharge rate, and superior power capability \cite{xiong2018battery,wang2017review,feng2019review}. They are extensively deployed in electric vehicles (EVs), portable electronics, grid-scale renewable energy storage systems, aerospace platforms, and emerging smart-grid infrastructures \cite{he2011state,berecibar2016critical,zheng2018state}. The safe and efficient operation of these systems depends critically on advanced battery management systems (BMS), which monitor and estimate key internal states including state of charge (SOC), state of health (SOH), and state of power (SOP).\cite{xiong2018battery,wang2017review,feng2019review}.

Among the various internal battery states, the state of charge (SOC) is one of the most fundamental and operationally significant quantities. From an electrochemical perspective, SOC reflects the normalized concentration of lithium ions stored within the active electrode materials and is therefore directly related to the degree of lithium intercalation and deintercalation. Equivalently, it represents the fraction of electrochemically available charge remaining relative to the fully charged state. Accurate SOC estimation is essential for preventing overcharge and deep discharge, maximizing battery utilization, extending service life, and enabling reliable driving-range prediction in EVs \cite{movassagh2021critical,piller2001methods,he2011state,zheng2018state}. Since SOC cannot be directly measured by any sensor, it must be inferred from externally accessible signals such as terminal voltage and applied current, making it an inherently challenging state estimation problem \cite{movassagh2021critical,piller2001methods}.

Numerous SOC estimation approaches have been proposed in the literature. Coulomb counting, which integrates the measured current over time, remains one of the simplest and most widely implemented methods because of its low computational burden \cite{piller2001methods,movassagh2021critical}. However, it suffers from cumulative integration error, sensor drift, uncertainty in Coulombic efficiency, and sensitivity to the initial SOC value \cite{movassagh2021critical,ng2011enhanced}. These limitations motivate the development of more sophisticated estimation frameworks.

Model-based approaches, particularly those employing equivalent circuit models (ECMs), have become widely adopted because they provide a computationally efficient representation of battery dynamics suitable for real-time implementation \cite{hu2012comparative,lai2018comparative,zhang2011review}. In these approaches, the battery is represented using combinations of resistors and capacitors that capture the instantaneous ohmic response and polarization dynamics \cite{hu2012comparative,barai2015identification}. Observer-based estimation algorithms such as the extended Kalman filter (EKF), unscented Kalman filter (UKF), sigma-point Kalman filter (SPKF), and adaptive Kalman filtering techniques are commonly combined with ECMs to estimate SOC in real time \cite{plett2004extended,plett2004unscented,plett2004sigma,sun2011adaptive,wei2022comparative}. Although these approaches often provide good estimation accuracy, their performance depends strongly on accurate identification of model parameters, including ohmic resistance, polarization resistances, and capacitances \cite{hu2012comparative,lai2018comparative,barai2015identification}. These parameters vary significantly with temperature, current rate, aging condition, and SOC, necessitating periodic recalibration and increasing modeling uncertainty \cite{wei2022comparative,lai2018comparative}.

Physics-based electrochemical models offer an alternative framework by explicitly describing lithium-ion transport, charge-transfer reactions, electrolyte diffusion, and solid-state diffusion within battery electrodes \cite{doyle1993modeling,newman2004modeling,santhanagopalan2006review}. Among these, the Doyle--Fuller--Newman (DFN) pseudo-two-dimensional model is widely regarded as the benchmark electrochemical model because of its high predictive fidelity \cite{doyle1993modeling,newman2004modeling}. Reduced-order electrochemical models have also been proposed to improve computational efficiency while retaining essential physical behavior \cite{smith2007control,santhanagopalan2006review}. Nevertheless, the complexity of these coupled partial differential equation systems often renders them computationally expensive for embedded real-time BMS applications \cite{smith2007control,wei2022comparative}.

More recently, data-driven approaches have attracted significant attention because they avoid explicit physical modeling and instead learn battery behavior directly from measurement data. Machine learning techniques including artificial neural networks, support vector machines, deep neural networks, and long short-term memory (LSTM) networks have demonstrated promising SOC estimation performance \cite{chemali2018long,li2019deep,yang2020state,sehgal2019battery}. Despite their predictive capabilities, these approaches generally require large quantities of labeled training data and often suffer from limited interpretability and reduced generalization across battery chemistries, operating conditions, and aging states \cite{chemali2018long,li2019deep,movassagh2021critical}.

To address these limitations, this work proposes a state-of-charge estimation framework based on Dynamic Mode Decomposition with control (DMDc). Dynamic Mode Decomposition (DMD), originally introduced by Schmid \cite{schmid2010dynamic}, is a data-driven technique that extracts coherent dynamical structures and spectral information directly from sequential measurements. Subsequent theoretical developments established deep connections between DMD and Koopman operator theory, providing a rigorous framework for representing nonlinear systems through linear operators acting on observables \cite{rowley2009spectral,mezic2005spectral,mezic2013analysis,tu2014dynamic}. Proctor et al.~\cite{proctor2016dynamic} extended classical DMD to systems influenced by external inputs, yielding DMDc, which identifies linear state-space models directly from input-output data without requiring prior knowledge of governing equations.

The theoretical foundation of DMDc is particularly attractive for battery applications because the applied current naturally acts as an external forcing input, while the measured terminal voltage serves as an observable of the underlying electrochemical dynamics. Furthermore, recent developments in Koopman theory have demonstrated that nonlinear dynamics can be represented approximately linearly in suitably chosen observable spaces \cite{mezic2013analysis,mauroy2020koopman,peitz2019koopman,surana2016koopman}. Time-delay embedding techniques based on Takens' embedding theorem provide a practical means of constructing such observable spaces directly from measured signals \cite{takens1981detecting}. The connection between Hankel delay embeddings, Koopman invariant subspaces, and DMD has been rigorously established in several recent studies \cite{arbabi2017ergodic,brunton2017chaos,kamb2020time,kutz2016dynamic,brunton2022data}.

In this work, time-delay (Hankel) embedding is applied to terminal voltage measurements obtained from hybrid pulse power characterization (HPPC) testing \cite{smith2010pulse,dubarry2010evaluation}. The resulting lifted state representation enables DMDc to identify the dominant battery dynamics directly from measured voltage and current data. Eigen-decomposition of the identified system matrix reveals a hierarchy of dynamic modes associated with intrinsic battery processes. Notably, the SOC dynamics emerge naturally as the mode whose eigenvalue lies closest to unity, consistent with the integrator structure implied by charge conservation \cite{plett2004extended,movassagh2021critical,labib2026modeling}.

A distinguishing advantage of the proposed framework is that it bypasses explicit parameter identification of equivalent circuit models while still recovering physically interpretable dynamical information. Consequently, the proposed approach provides a computationally efficient pathway toward data-driven SOC estimation. The remainder of this paper is organized as follows. 
Section~\ref{sec:theory} presents the theoretical framework, including the equivalent circuit model, Extended Kalman Filter formulation, Hankel time-delay embedding, and Dynamic Mode Decomposition with control (DMDc), along with the experimental methodology and SOC initialization procedure. Section~\ref{sec:results} presents the voltage reconstruction results, Koopman spectral interpretation of battery dynamics, and comparative SOC estimation performance analysis. The limitations of the present approach and future research directions are also discussed. Finally, Section~\ref{sec:conclusion} summarizes the major findings and contributions of this work.

\section{Methodology}\label{sec:theory}

\subsection{Battery Voltage Dynamics as a Forced Nonlinear System}

A lithium-ion battery can be regarded as a forced dynamical system in which the terminal voltage $V_t$ evolves under the influence of an external input, the applied current $I$:
\begin{equation}
\frac{dV_t}{dt}=f(V_t,I).
\label{eq:nonlinear}
\end{equation}
Because the relationship between terminal voltage, internal electrochemical states, and applied current is strongly nonlinear, $f(\cdot)$ in Eq.~\ref{eq:nonlinear} cannot, in general, be reduced to a linear differential equation. To enable a linear data-driven treatment of this nonlinear system, the voltage and current measurements are lifted into a higher-dimensional observable space in which the dynamics behave approximately linearly. This is achieved through Hankel time-delay embedding, a procedure that is closely connected to Koopman operator theory \cite{brunton2017chaos,kamb2020time}.

\subsection{Hankel Time-Delay Embedding}

Time-delay embedding reconstructs a higher-dimensional state space from a scalar time series by stacking successive delayed measurements into a Hankel matrix. Given a discrete voltage sequence $[v(1),v(2),\dots,v(N)]$ sampled at $N$ time points with embedding dimension $d$, the Hankel matrix is constructed as
\begin{equation}
H_v=
\begin{bmatrix}
v(1) & v(2) & \cdots & v(n) \\
v(2) & v(3) & \cdots & v(n+1) \\
\vdots & \vdots & \ddots & \vdots \\
v(d) & v(d+1) & \cdots & v(N)
\end{bmatrix},
\label{eq:hankel}
\end{equation}
where $n=N-d+1$ and $H_v \in \mathbb{R}^{d\times n}$. This procedure raises the dimensionality of the original scalar voltage signal from one to $d$. The additional rows of $H_v$ encode temporal correlations that are otherwise inaccessible from a single instantaneous voltage measurement, allowing the stacked observable to capture hidden internal states of the battery, such as the state of charge (SOC) and polarization voltages, that are not directly measurable \cite{brunton2017chaos}.

\subsection{Dynamic Mode Decomposition with Control}

Once the Hankel matrix of voltage is formed, Dynamic Mode Decomposition with control (DMDc) \cite{proctor2016dynamic}, an extension of the original DMD algorithm \cite{schmid2010dynamic}, is applied to identify a linear state-space model relating the lifted voltage states to the applied current input. The embedded state is assumed to evolve as
\begin{equation}
x_{k+1}=Ax_k+Bu_k,
\label{eq:dmdc_model}
\end{equation}
where $x_k$ is the $k^{\mathrm{th}}$ column of the Hankel-embedded voltage state, $u_k$ is the corresponding applied current, $A$ is the state-transition matrix, and $B$ is the input matrix. Collecting snapshots into the data matrices $X=[x_1,\dots,x_{n-1}]$, $X'=[x_2,\dots,x_n]$, and $U=[u_1,\dots,u_{n-1}]$, the relationship in Eq.~\ref{eq:dmdc_model} can be written as $X'\approx AX+BU$. Stacking $X$ and $U$ into the augmented matrix $\Omega=[X\;\,U]^{T}$ and defining $G=[A\;\,B]$, the best-fit operators are obtained via the least-squares solution
\begin{equation}
G \approx X'\,\Omega^{\dagger},
\label{eq:dmdc_ls}
\end{equation}
where $\Omega^{\dagger}$ denotes the Moore--Penrose pseudoinverse, computed using a truncated singular value decomposition for numerical robustness. The resulting matrix $A$ governs the intrinsic voltage dynamics, while $B$ captures the influence of the applied current.

\subsection{Eigen-Decomposition and Mode Identification}

The system matrix $A$ identified from Eq.~\ref{eq:dmdc_ls} captures the full dynamics of the lifted voltage state. To extract the internal dynamical structure, eigen-decomposition is applied to $A$,
\begin{equation}
A\Phi=\Phi\Lambda,
\label{eq:eig}
\end{equation}
where $\Lambda=\mathrm{diag}(\lambda_1,\dots,\lambda_r)$ contains the discrete-time eigenvalues and $\Phi=[\phi_1,\dots,\phi_r]$ contains the corresponding eigenvectors, or dynamic modes. Each eigenvalue $\lambda_i$ governs the temporal evolution of its associated mode, so that the full set of eigenvalues collectively represents the time motion of the distinct internal states and dynamical processes contributing to the measured voltage.

\subsection{SOC Mode Identification}

The continuous-time evolution of SOC under Coulomb counting is given by
\begin{equation}
z(t)=z_0-\frac{1}{Q}\int_0^t \eta\, i(\tau)\,d\tau,
\label{eq:soc_cc}
\end{equation}
where $Q$ is the cell capacity and $\eta$ is the Coulombic efficiency. Taking the Laplace transform of Eq.~\ref{eq:soc_cc} and neglecting the initial-condition term yields the transfer function
\begin{equation}
\frac{Z(s)}{I(s)}=-\frac{\eta}{Qs}.
\label{eq:soc_tf}
\end{equation}
The transfer function in Eq.~\ref{eq:soc_tf} has a pole at the origin ($s=0$), which corresponds to a continuous-time eigenvalue $\lambda_c=0$. Continuous-time and discrete-time eigenvalues are related through the standard sampling relation
\begin{equation}
\lambda_d=e^{\lambda_c\, \Delta t},
\label{eq:eig_map}
\end{equation}
where $\Delta t$ is the sampling interval. Substituting $\lambda_c=0$ into Eq.~\ref{eq:eig_map} gives $\lambda_d=1$. The SOC dynamics therefore correspond, in discrete time, to a Koopman eigenvalue equal to unity, reflecting the marginally stable, integrator-type nature of charge conservation.

% \subsection*{Decoupling Terminal Voltage Dynamics}

% The measured terminal voltage results from the superposition of three physically distinct contributions: the open-circuit voltage (OCV), the RC polarization overpotential $V_p$, and the instantaneous ohmic drop,
% %
% \begin{equation}
% V_t = OCV + V_p + IR_0.
% \label{eq:vt_decomp}
% \end{equation}
% %
% Each contribution in Eq.~\ref{eq:vt_decomp} evolves on a distinct timescale. The ohmic drop $IR_0$ responds essentially instantaneously to the applied current. The polarization voltage $V_p$ exhibits non-oscillatory exponential relaxation associated with charge-transfer and diffusion processes. The OCV, being a monotonic function of SOC, evolves on a much slower timescale governed by the integrator dynamics derived above. Because these three contributions occupy well-separated regions of the eigenspectrum, the eigen-decomposition of $A$ cleanly separates their respective time motions.

Consistent with Eq.~\ref{eq:eig_map}, the slow, non-oscillatory SOC dynamics correspond to a discrete-time eigenvalue $\lambda_d=1$, equivalent to a continuous-time eigenvalue $\lambda_c=0$. The OCV, being a monotonic function of SOC, evolves on a much slower timescale governed by the integrator dynamics derived above. Since this eigenvalue has no imaginary component, it is consistent with the non-oscillatory character of the OCV mode; and since its continuous-time real part is also zero, the corresponding dynamics neither grow nor decay, but instead evolve quasi-statically.

To decouple the contribution of each internal dynamical process from the measured terminal voltage, the voltage Hankel matrix is projected onto the identified mode matrix,
\begin{equation}
Y=\Phi^{-1}H_v.
\label{eq:projection}
\end{equation}
The row of $Y$ corresponding to the SOC eigenvalue ($\lambda_d\approx 1$) yields the raw SOC-sensitive modal coordinate. After Min--Max Normalization of this raw coordinate over the operating SOC range, the resulting signal provides the DMDc-based SOC estimate.

\begin{figure}[!t]
\centering
\includegraphics[width=0.95\textwidth,
% height=0.5\textheight,
keepaspectratio]{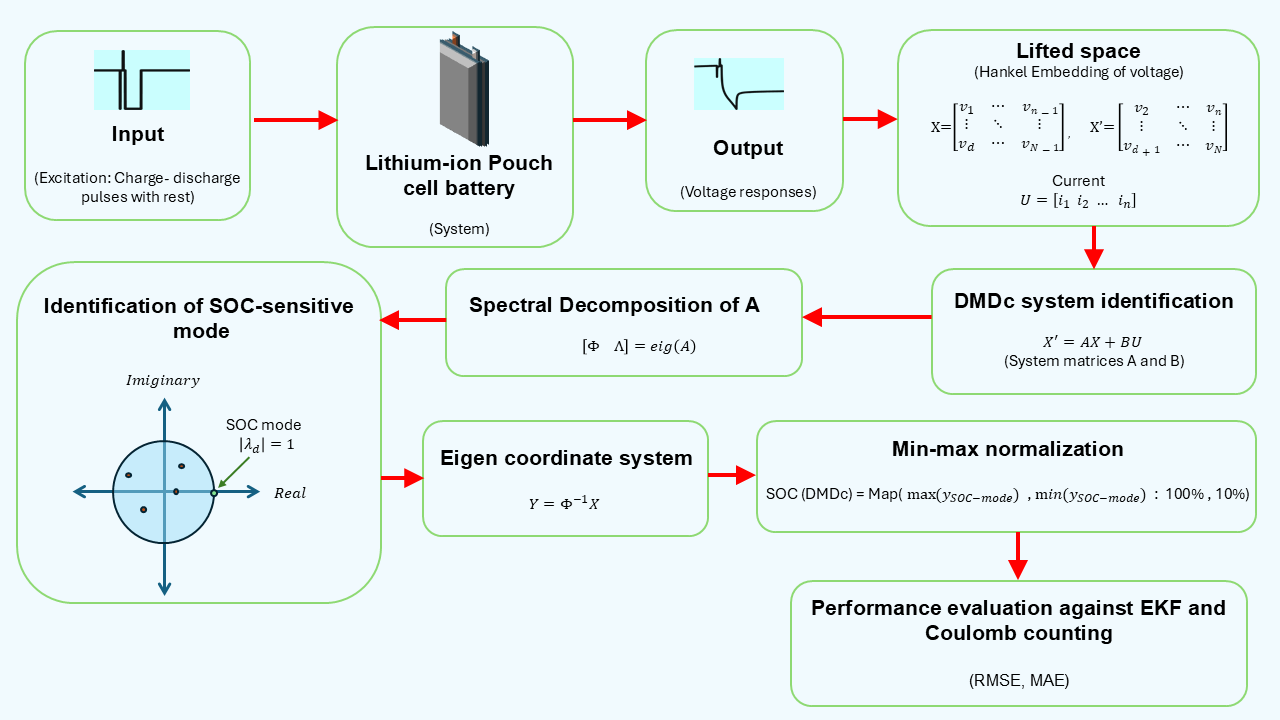}
\caption{
Workflow of the proposed DMDc-based SOC estimation framework.
HPPC current and voltage measurements are first arranged using
Hankel time-delay embedding. DMDc is then applied to identify
the lifted linear operator. Koopman eigenvalues, modes, and modal
coordinates are subsequently extracted through spectral decomposition.
The modal coordinate associated with the eigenvalue closest to the
unit circle is identified as the SOC-sensitive observable and used for
SOC estimation.
}
\label{fig:flowchart}
\end{figure}

\subsection{State-Space Battery Model and Extended Kalman Filter}

An Extended Kalman Filter (EKF) based on the 2RC ECM was implemented as a benchmark. The battery state vector is defined as
\begin{equation}
x_k=
\begin{bmatrix}
z_k & V_{1,k} & V_{2,k}
\end{bmatrix}^{T},
\end{equation}
where $z_k$ is the SOC and $V_{1,k}$, $V_{2,k}$ are the polarization voltages of the two RC branches. The state equations governing their discrete-time evolution are
\begin{align}
z_{k+1} &= z_k - \frac{\eta I_k \Delta t}{Q}, \\
V_{j,k+1} &= e^{-\Delta t / \tau_j} V_{j,k} + R_j\!\left(1 - e^{-\Delta t / \tau_j}\right) I_k, \quad j = 1,2,
\end{align}
where $Q$ is the nominal capacity, $\eta$ is the Coulombic efficiency, $\Delta t$ is the sampling interval, and $\tau_j = R_j C_j$ is the time constant of the $j^{\mathrm{th}}$ RC branch. The terminal voltage measurement equation is
\begin{equation}
V_{t,k} = OCV(z_k) - I_k R_0 - V_{1,k} - V_{2,k},
\label{eq:meas}
\end{equation}
where $OCV(z_k)$ is the nonlinear open-circuit voltage function and $R_0$ is the ohmic resistance. Because Eq.~\ref{eq:meas} is nonlinear in $z_k$, the EKF linearizes the measurement model about the current state estimate via the Jacobian $\partial OCV / \partial z$ at each correction step. The filter then alternates between propagating the state forward using the current input and correcting the estimate using the voltage innovation, weighted by the Kalman gain.

%------------------------------------------------------------------------------------------------------------
%
\section{Results and Discussion}\label{sec:results}

\subsection{Experimental Data}

The HPPC test was employed to generate a rich excitation profile containing charge pulses, discharge pulses, and relaxation intervals, thereby enabling the concurrent observation of both transient and long-timescale battery dynamics. The applied current and measured voltage profiles are presented in Fig.~\ref{fig:HPPC}. These experimental measurements were subsequently utilized for equivalent circuit model (ECM) parameter identification, extended Kalman filter (EKF) implementation, and dynamic mode decomposition with control (DMDc)-based data-driven modeling.

\subsection{ECM Parameter Variation and EKF Implementation}

The identified 2RC ECM parameters exhibit substantial variation across different state-of-charge (SOC) regions, indicating the strongly nonlinear and operating-condition-dependent nature of lithium-ion battery dynamics. In particular, the polarization resistances and capacitances do not remain constant over the SOC range; rather, their values are significantly influenced by the prevailing battery operating condition. As a result, such variability introduces non-negligible modeling uncertainty into ECM-based estimators and necessitates repeated parameter identification procedures for different operating conditions, which in turn limits the practical applicability of purely model-based approaches.

In order to provide a rigorous comparison with the proposed data-driven framework, an EKF based on the identified 2RC ECM was implemented. The EKF utilized the following initial covariance and noise covariance matrices:

\begin{equation}
\begin{aligned}
P_{\mathrm{init}} &= \mathrm{diag}\!\left(0.05^2,\;0.01^2,\;0.01^2\right), \\
Q_{\mathrm{noise}} &= \mathrm{diag}\!\left(10^{-6},\;10^{-3},\;10^{-3}\right), \\
R_{\mathrm{noise}} &= 10^{-4}.
\end{aligned}
\label{eq:ekf_covariances}
\end{equation}

It should be noted that the initial covariance matrix $P_{\mathrm{init}}$ reflects comparatively larger uncertainty in the initial SOC estimate relative to the polarization voltages. The process-noise covariance $Q_{\mathrm{noise}}$ is employed to account for model uncertainty associated with unmodeled battery dynamics and parameter variations across operating conditions, whereas the measurement-noise covariance $R_{\mathrm{noise}}$ reflects the inherent uncertainty in the voltage sensor measurements.

\subsection{Voltage Reconstruction Accuracy}

\begin{figure}[!t]
\centering

\begin{subfigure}{0.51\textwidth}
    \raggedright
    \textbf{\textit{A}}\; HPPC test voltage and current profile\\[3pt]
    \includegraphics[width=\linewidth]{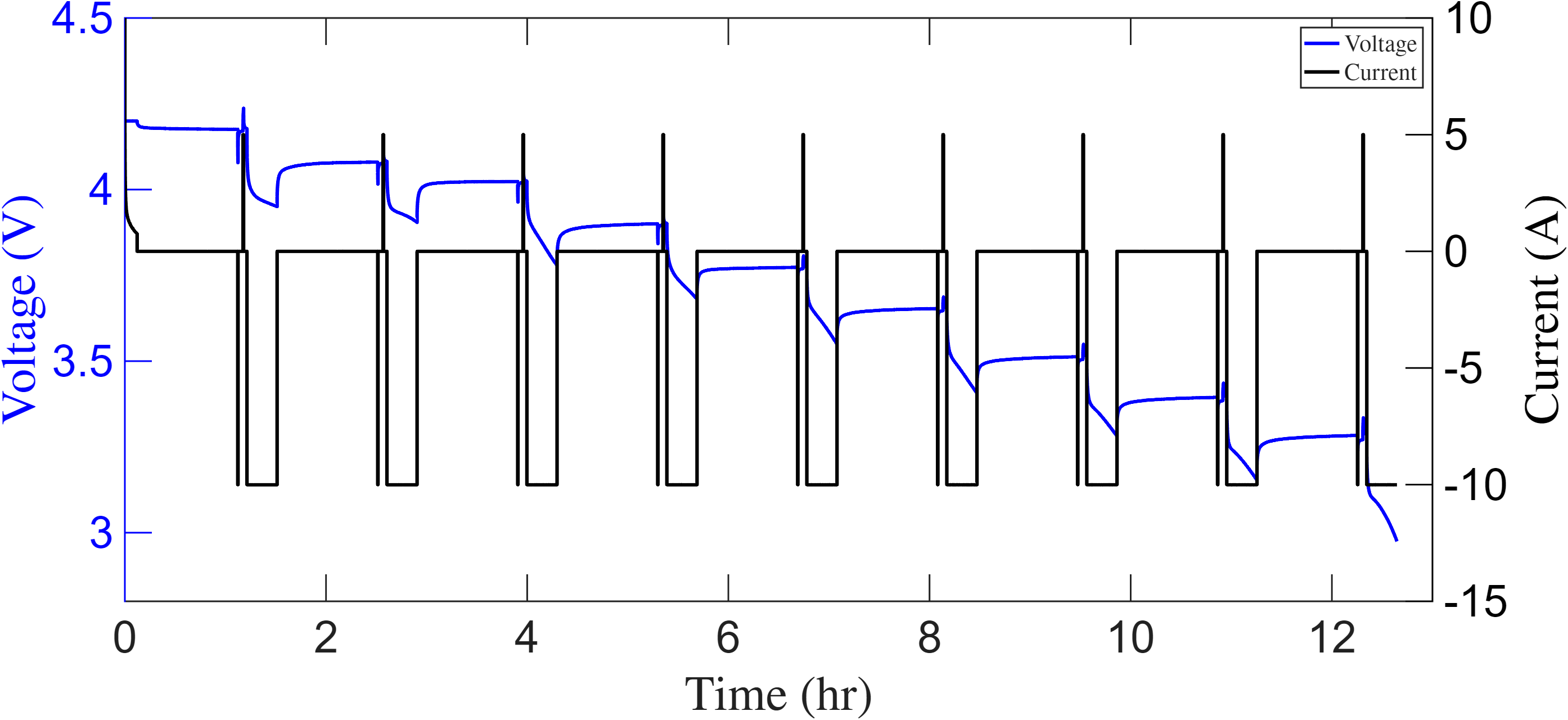}
    \phantomcaption
    \label{fig:HPPC}
\end{subfigure}
\hfill
\begin{subfigure}{0.47\textwidth}
    \raggedright
    \textbf{\textit{B}}\; Measured vs.\ predicted terminal voltage\\[3pt]
    \includegraphics[width=\linewidth]{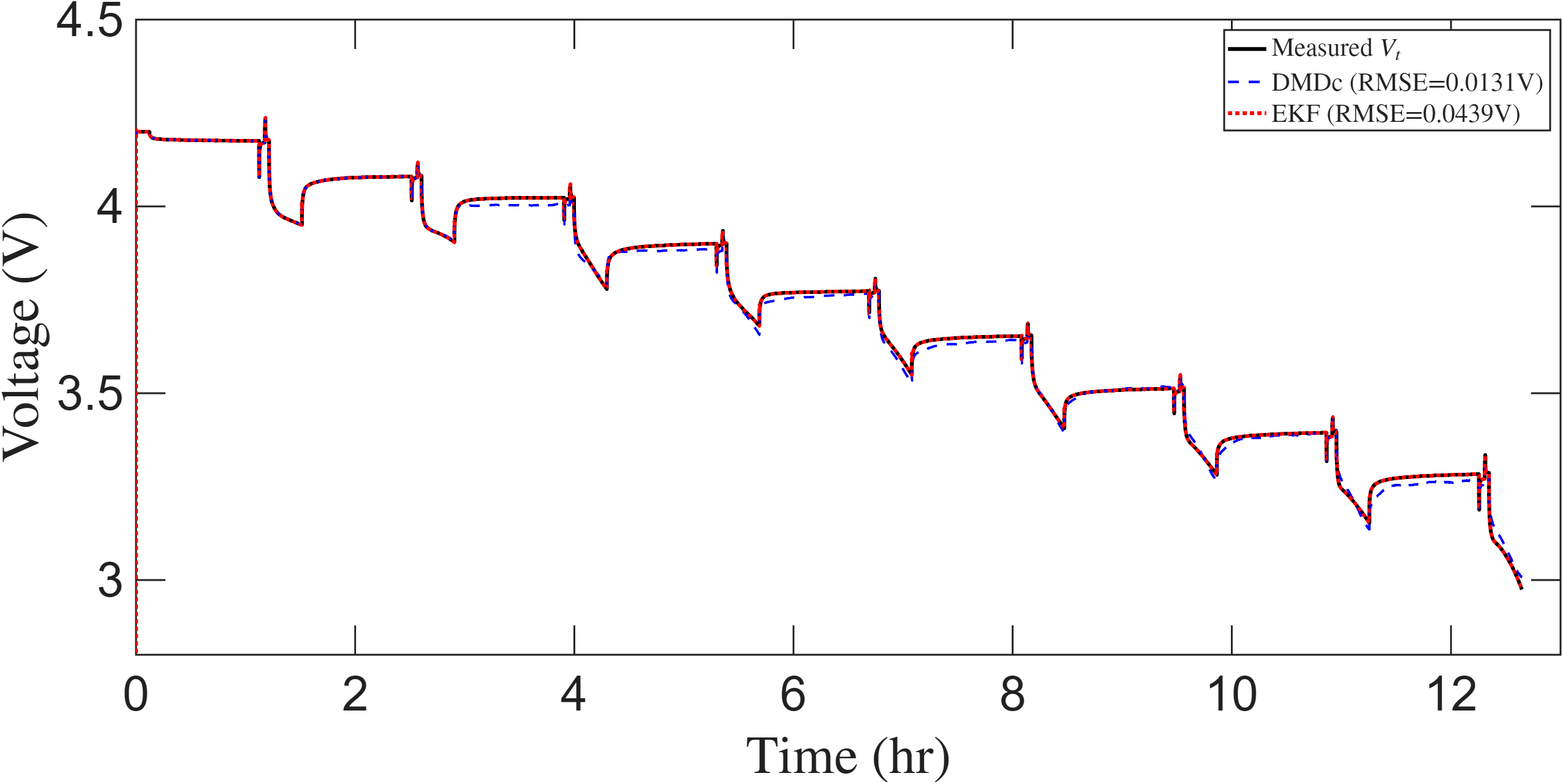}
    \phantomcaption
    \label{fig:Vpred}
\end{subfigure}

\vspace{0.5cm}

\begin{subfigure}{0.47\textwidth}
    \raggedright
    \textbf{\textit{C}}\; Zoomed view of voltage transient region\\[3pt]
    \includegraphics[width=\linewidth]{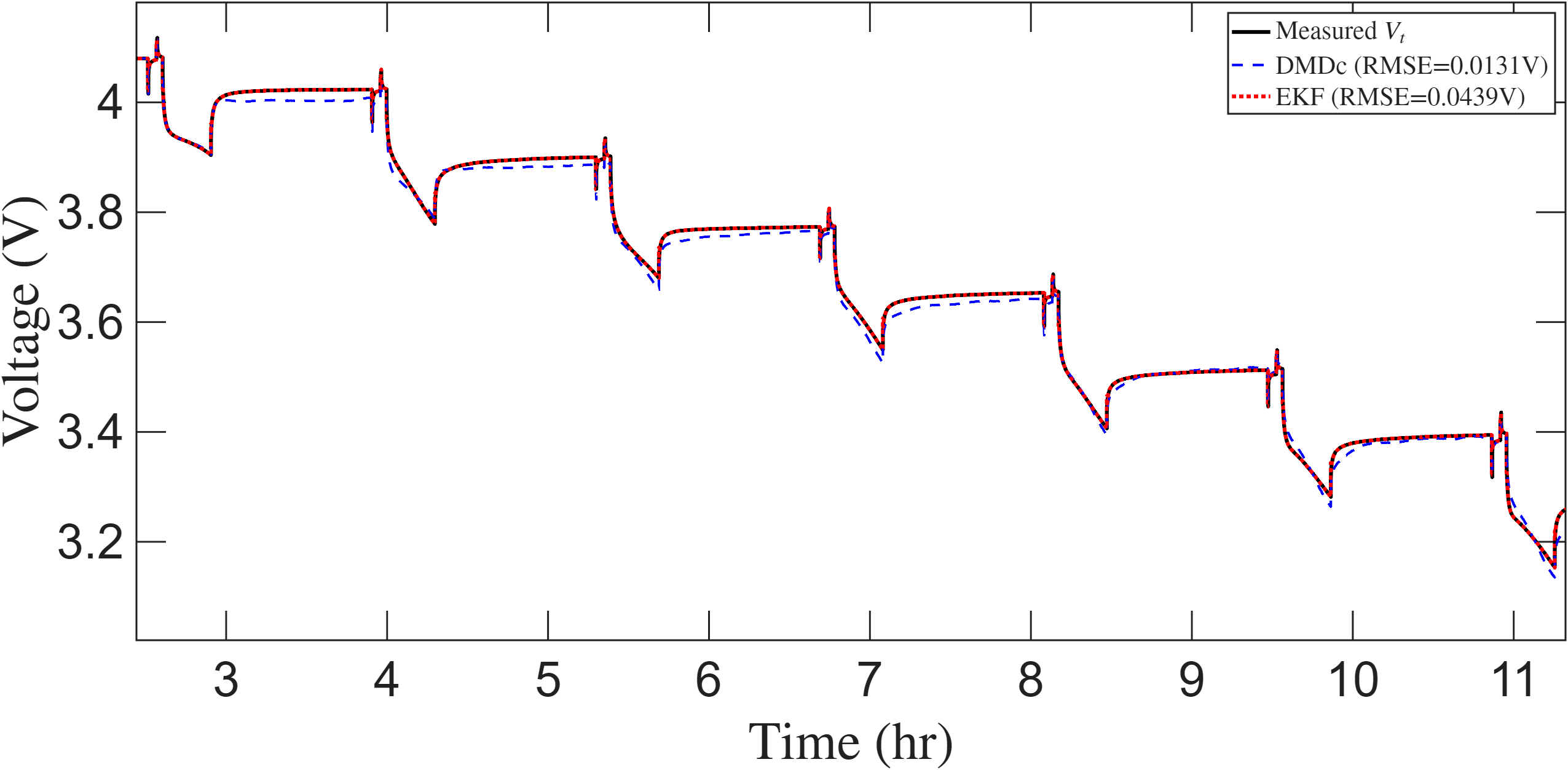}
    \phantomcaption
    \label{fig:VpredZoom}
\end{subfigure}
\hfill
\begin{subfigure}{0.47\textwidth}
    \raggedright
    \textbf{\textit{D}}\; Pointwise voltage prediction error\\[3pt]
    \includegraphics[width=\linewidth]{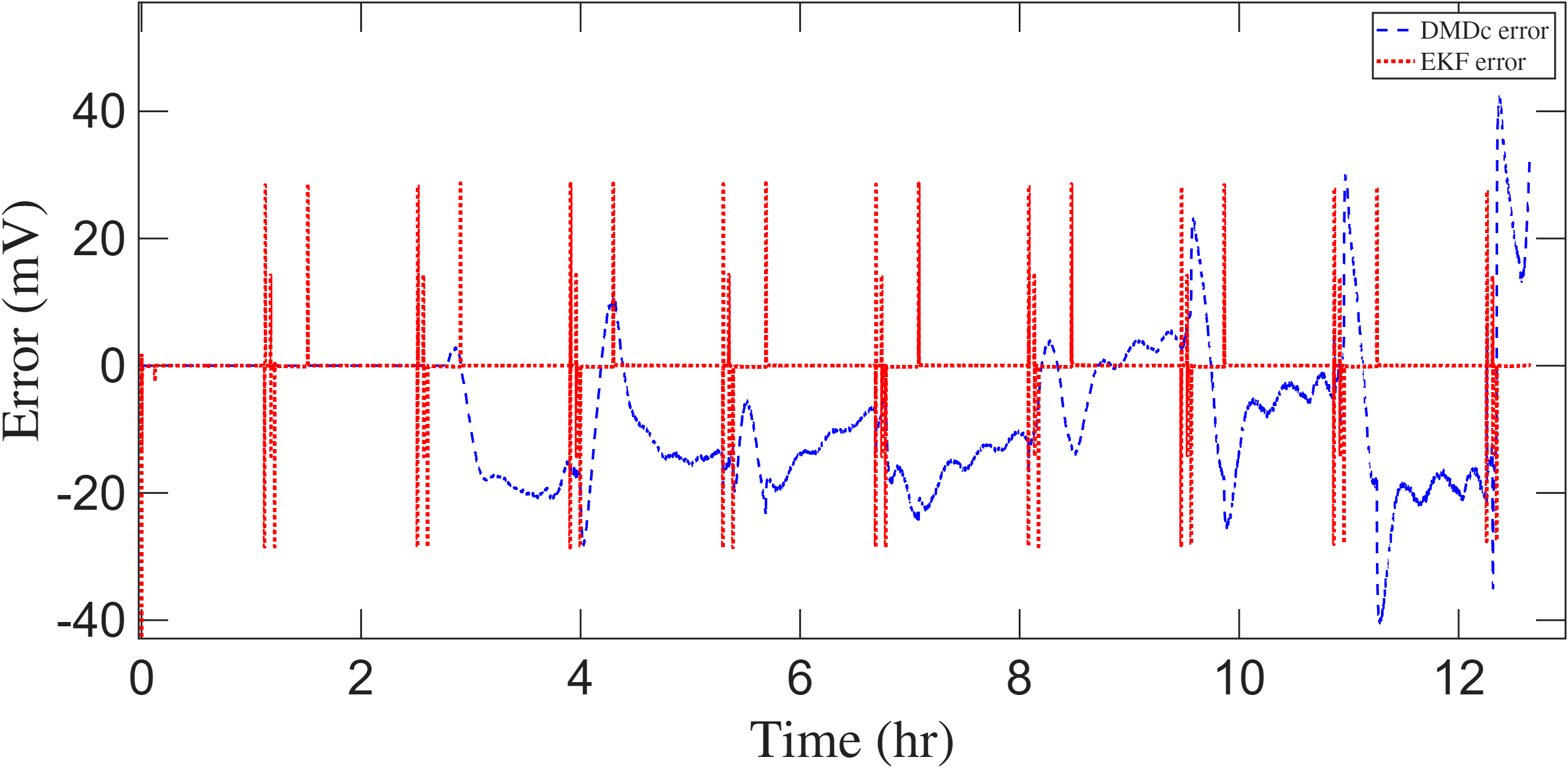}
    \phantomcaption
    \label{fig:Verror}
\end{subfigure}

\caption{DMDc-based terminal voltage prediction evaluated on the full HPPC test cycle.
(\textit{A}) Hybrid Pulse Power Characterization (HPPC) test profile showing the
measured terminal voltage and applied current excitation, including
charge-sustaining rest periods and symmetric charge/discharge pulses across
the full SOC range.
(\textit{B}) Comparison between experimentally measured terminal voltage and the
DMDc-predicted response over the complete test cycle. The DMDc model,
driven solely by the measured current input, accurately reconstructs both
the quasi-static OCV trajectory and the fast pulse-induced voltage transients.
(\textit{C}) Zoomed view of a representative segment of the voltage response,
highlighting the model's ability to capture sharp voltage transitions during
high-rate current pulses with minimal phase and amplitude error.
(\textit{D}) Pointwise voltage prediction error across the full HPPC test,
demonstrating bounded residuals throughout the cycle and confirming the
generalization capability of the Hankel-DMDc framework beyond the
training window.}
\label{fig:Exp-volt}
\end{figure}

In the context of the proposed framework, the DMDc model was trained using approximately 75\% of the measured voltage and current data. A Hankel time-delay embedding dimension of 2000 was employed in order to reconstruct the lifted observable space and to retain sufficient temporal information associated with the internal electrochemical dynamics. It can be observed from Fig.~\ref{fig:Vpred} that the identified DMDc model demonstrates strong voltage reconstruction capability over the complete test cycle. The predicted response accurately captures both the rapid transient voltage variations induced by high-rate current pulses and the slow relaxation behavior observed during the rest intervals. It should be mentioned here that the Hankel-DMDc framework was evaluated exclusively on the held-out portion of the data, thereby confirming its generalization capability beyond the training window. Fig.~\ref{fig:Vpred} shows the comparison between the experimentally measured terminal voltage, EKF and the DMDc-predicted response over the complete test cycle. Fig.~\ref{fig:VpredZoom} provides a zoomed view of a representative transient segment, confirming accurate phase and amplitude tracking during high-rate current pulses. 

The pointwise voltage prediction error across the full HPPC test cycle is shown in Fig.~\ref{fig:Verror}. The residuals remain bounded throughout the cycle, including during high-rate pulse transients. Furthermore, it can be observed from Table~\ref{tab:voltage_summary} that the DMDc framework achieves a substantially lower voltage prediction RMSE of 0.0131~V compared to the EKF value of 0.0439~V, indicating that the delay-embedded Koopman representation more effectively captures both transient and long-timescale battery dynamics directly from measurement data, without reliance on a parameterized physics model.
\begin{table}[H]
\centering
\caption{Voltage prediction accuracy comparison}
\label{tab:voltage_summary}
\begin{tabular}{lcc}
\hline
Method & RMSE (V) & MAE (V) \\
\hline
DMDc & 0.0131 & 0.0100 \\
EKF  & 0.0439 & 0.0007 \\
\hline
\end{tabular}
\end{table}

\subsection{Koopman Spectral Interpretation of Battery Dynamics}

Following DMDc identification, eigen-decomposition of the learned state-transition matrix was performed, revealing a collection of dynamic modes associated with distinct temporal behaviors of the battery system. Among the identified eigenvalues, the mode whose eigenvalue lies closest to the unit circle was identified as the SOC-sensitive Koopman mode; this spectral identification constitutes the foundation of the proposed estimation strategy.

It can be observed from Fig.~\ref{fig:unit-circle} that, from the 2000 identified modes, the dominant SOC-associated mode corresponds to mode number 457 with eigenvalue $\lambda_{\mathrm{SOC}}=1.000001$. This eigenvalue lies approximately on the unit circle with nearly zero oscillatory frequency, which is consistent with the integrator-type dynamics of SOC derived from charge conservation. Since SOC evolves through cumulative charge accumulation rather than exponential decay, its corresponding Koopman dynamics remain marginally stable over time. Due to lifting of voltage, some spurious modes also evolve. 
% Some modes, characterized by eigenvalues with decay rates further from unity, correspond to transient polarization and diffusion processes analogous to the RC branches of the equivalent circuit model.

This spectral interpretation demonstrates that the DMDc framework naturally separates battery dynamics according to their intrinsic temporal scales without requiring explicit electrochemical parameterization, which constitutes a significant advantage over physics-based estimation approaches.

\subsection{SOC Estimation Performance}

\begin{figure}[!t]
\centering

\begin{subfigure}{0.47\textwidth}
    \raggedright
    \textbf{\textit{A}}\; Eigenvalue spectrum in the complex plane\\[3pt]
    \includegraphics[width=\linewidth]{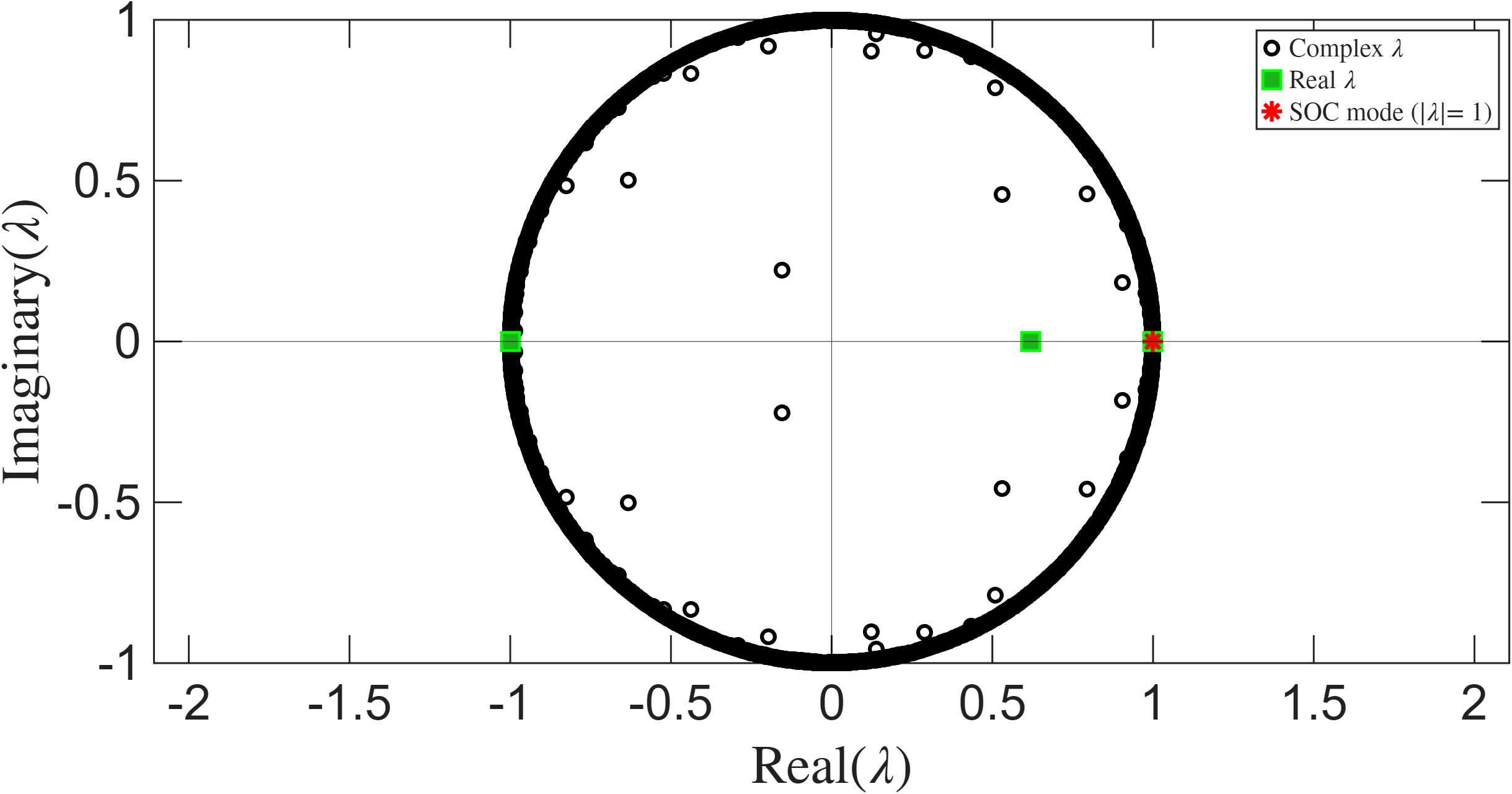}
    \phantomcaption
    \label{fig:unit-circle}
\end{subfigure}
\hfill
\begin{subfigure}{0.49\textwidth}
    \raggedright
    \textbf{\textit{B}}\; SOC estimation across full HPPC cycle\\[3pt]
    \includegraphics[width=\linewidth]{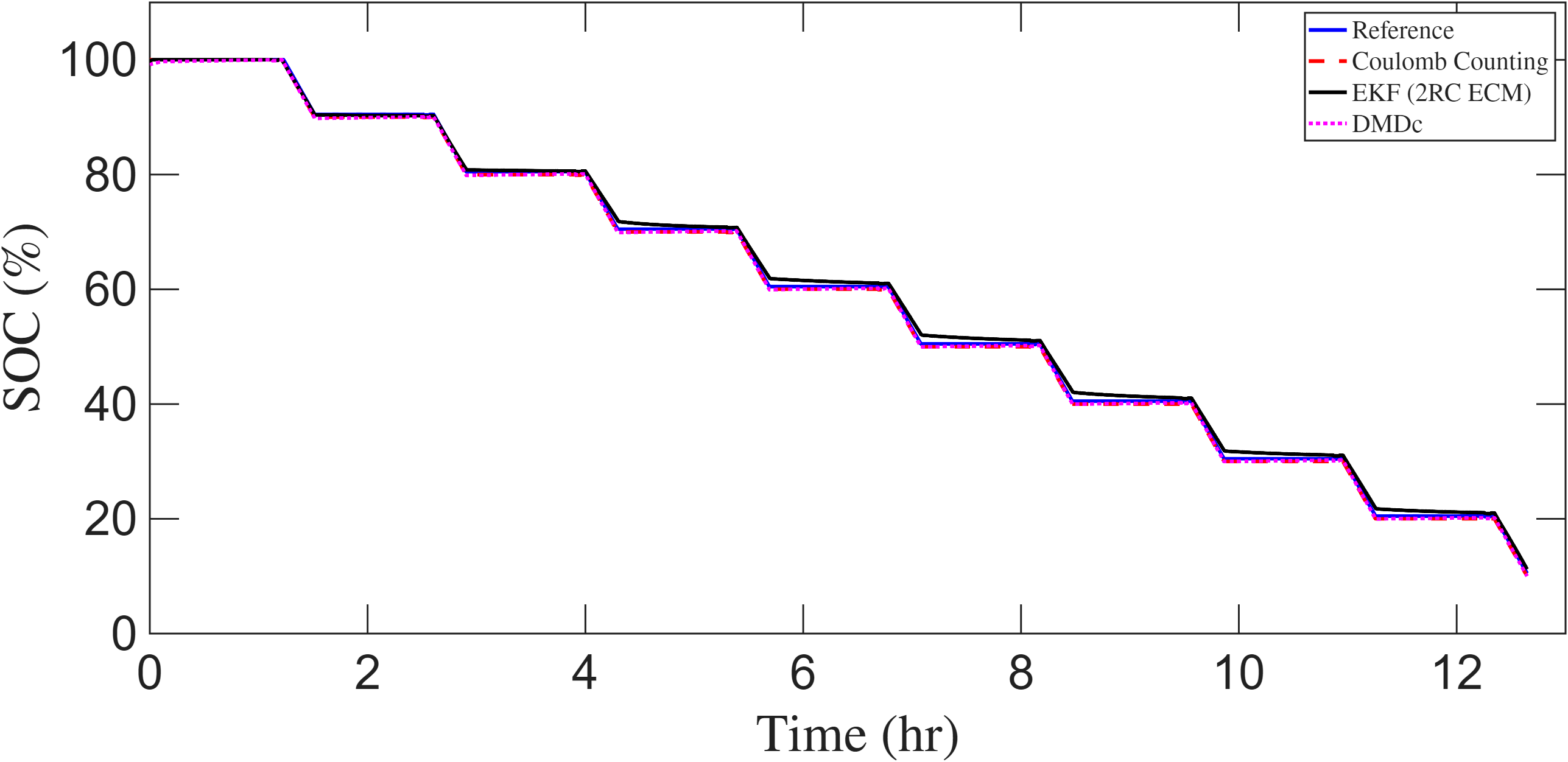}
    \phantomcaption
    \label{fig:soc}
\end{subfigure}

\vspace{0.5cm}

\begin{subfigure}{0.47\textwidth}
    \raggedright
    \textbf{\textit{C}}\; Zoomed view of SOC transient region\\[3pt]
    \includegraphics[width=\linewidth]{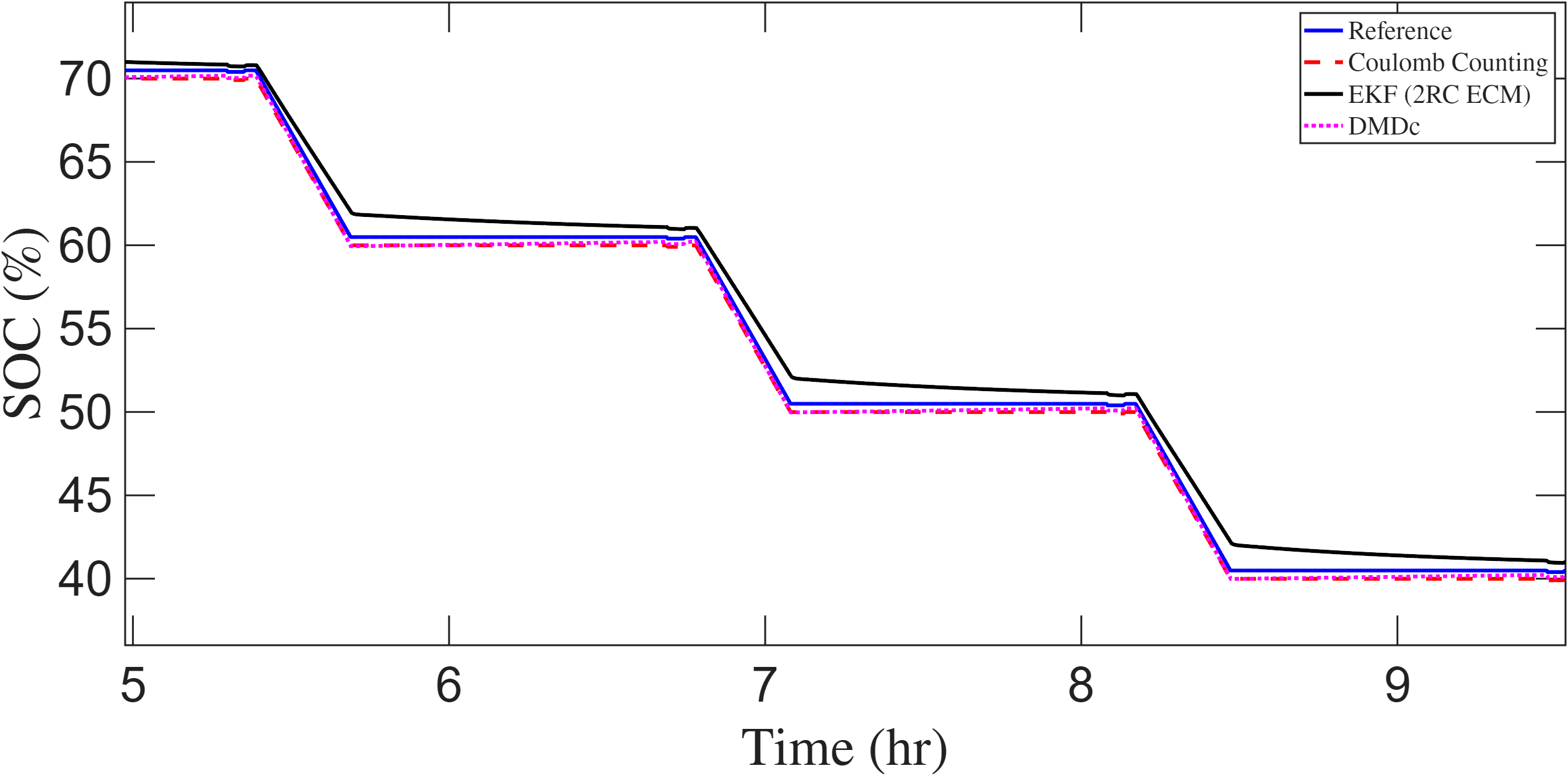}
    \phantomcaption
    \label{fig:soc-zoom}
\end{subfigure}
\hfill
\begin{subfigure}{0.51\textwidth}
    \raggedright
    \textbf{\textit{D}}\; Pointwise SOC estimation error\\[3pt]
    \includegraphics[width=\linewidth]{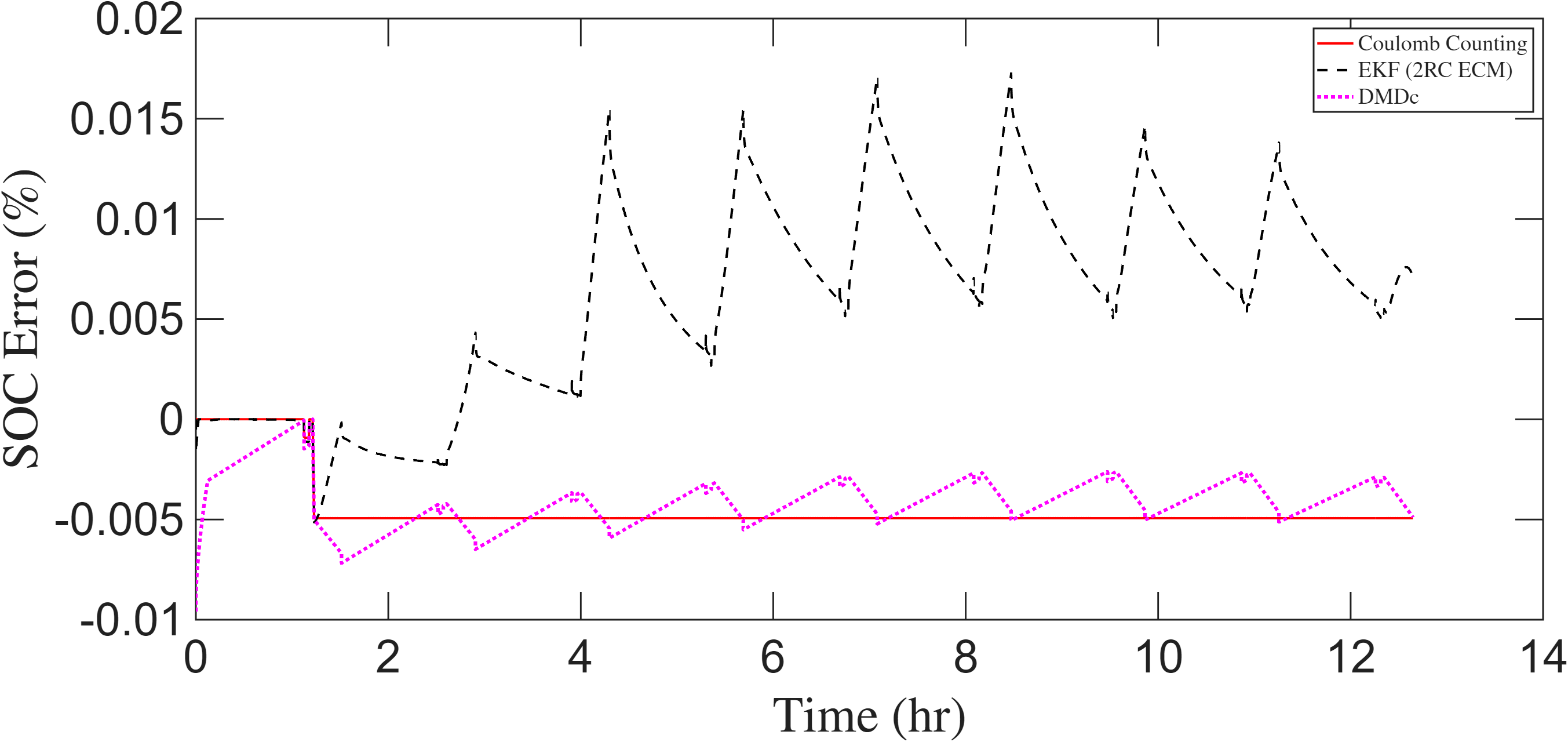}
    \phantomcaption
    \label{fig:soc-error}
\end{subfigure}

\caption{Koopman-theoretic analysis of Li-ion battery dynamics via DMDc.
(\textit{A}) Eigenvalues of the learned DMDc operator plotted in the complex plane
against the unit circle. Eigenvalues near the unit circle correspond to slowly
decaying dynamic modes; the SOC-sensitive mode (highlighted) exhibits the
smallest decay rate, confirming its role in governing long-timescale capacity
fade behavior.
(\textit{B}) Comparison of DMDc-based SOC estimation against Coulomb-counting SOC
and the reference SOC derived from experimental discharge capacity data across
the full HPPC test cycle. The DMDc estimator accurately tracks both slow
discharge trends and fast pulse-induced transients.
(\textit{C}) Zoomed view of a representative SOC segment highlighting transient
tracking fidelity during high-rate current pulses, where Coulomb-counting
accumulates drift and the reference SOC exhibits stepwise changes.
(\textit{D}) Pointwise SOC estimation error for all three methods, demonstrating
that the DMDc-based approach maintains consistently lower error magnitude
with bounded deviation from the experimental reference.}
\label{fig:koopman_results}
\end{figure}

The modal coordinate associated with the slowest Koopman mode was extracted and interpreted as the SOC-sensitive observable. After min--max normalization, the extracted modal coordinate provides the DMDc-based SOC estimate. 
% The comparison between DMDc-based SOC estimation, EKF, Coulomb-counting SOC, and reference SOC is presented in Fig.~\ref{fig:soc}--\ref{fig:soc-error}.
Fig.~\ref{fig:soc} presents the comparison of DMDc-based SOC estimation against EKF, Coulomb-counting estimation, and the reference SOC derived from experimental discharge capacity data across the full HPPC cycle. The DMDc estimator tracks both the slow discharge trend and the fast pulse-induced transients accurately. Fig.~\ref{fig:soc-zoom} shows a zoomed view of a representative SOC segment during high-rate current pulses, where Coulomb counting begins to accumulate drift while the DMDc estimate remains closely aligned with the reference. Fig.~\ref{fig:soc-error} presents the pointwise SOC estimation error for all three methods across the full cycle.
The SOC estimation accuracy was evaluated using the root-mean-square error (RMSE) and mean absolute error (MAE) computed relative to the reference SOC. The results are summarized in Tables~\ref{tab:soc_summary}.

\begin{table}[H]
\centering
\caption{SOC estimation performance comparison}
\label{tab:soc_summary}
\begin{tabular}{lcc}
\hline
Method & RMSE (\%) & MAE (\%) \\
\hline
Coulomb counting  & 0.0047 & 0.0045 \\
EKF (2RC ECM)     & 0.0078 & 0.0065 \\
DMDc              & 0.0043 & 0.0041 \\
\hline
\end{tabular}
\end{table}

It can be observed from Table~\ref{tab:soc_summary} that the DMDc-based approach achieves the lowest SOC estimation error among all compared methods, with an RMSE of 0.0043\% and an MAE of 0.0041\%. Although Coulomb counting provides comparatively accurate short-duration SOC estimation, it remains sensitive to cumulative integration drift and initialization uncertainty, which in turn limits its reliability over extended operating cycles. The EKF performance is additionally influenced by parameter mismatch and model uncertainty arising from variations in ECM parameters across different SOC regions, as previously discussed. 

%-------------------------------------------------------------------------
%-------------------------------------------------------------------------

\section{Conclusion}\label{sec:conclusion}

The proposed framework demonstrates that Koopman spectral analysis provides a physically meaningful decomposition of lithium-ion battery dynamics without requiring explicit electrochemical parameterization. In this context, it is of paramount importance to note that the marginally stable mode with eigenvalue closest to unity directly encodes charge-conservation dynamics. As a result, the Hankel-DMDc formulation effectively achieves a data-driven separation of dynamic timescales that is physically interpretable and consistent with established electrochemical theory.

Although the proposed framework successfully identifies physically interpretable battery dynamics and demonstrates accurate SOC trend estimation, some limitations remain and may be addressed in future investigations. In the first instance, it should be mentioned that the extracted SOC-sensitive modal coordinate does not directly provide quantitatively calibrated SOC values in physical units; min--max normalization is required to map the modal evolution into the SOC operating range. A more rigorous quantitative mapping between Koopman coordinates and physically measurable SOC constitutes an open problem that warrants further investigation.

In addition, the present DMDc formulation assumes a globally linear approximation within the lifted observable space. However, lithium-ion batteries exhibit strongly nonlinear behavior influenced by SOC and temperature variation, current rate, hysteresis, and degradation level. As a result, significant battery aging may alter the underlying electrochemical dynamics and shift the Koopman spectral structure, potentially reducing some estimation accuracy if the model is not updated or retrained accordingly. Furthermore, it should also be noted that multiple electrochemical processes may contribute jointly to a single Koopman mode — a phenomenon commonly referred to as mode mixing — which limits direct one-to-one correspondence between DMDc modes and exact electrochemical states.

Upon consideration of these factors, future research may focus on advanced Koopman-based decomposition techniques in order to improve spectral separation of closely coupled electrochemical processes. Nonlinear observable lifting strategies and adaptive Koopman representations capable of maintaining robustness under battery aging and temperature variation can also be investigated. Extension toward state-of-health (SOH) estimation and degradation diagnostics represents another important direction, since long-term spectral changes in Koopman eigenvalues may provide valuable indicators of capacity fade and internal resistance growth. Finally, real-time implementation within embedded battery management systems (BMS), and extension to different battery chemistries, battery packs, and other electrochemical energy-storage technologies, may be pursued. This work paves the way for the development of a thorough Koopman-theoretic framework for comprehensive battery state estimation and health monitoring.

\section*{Acknowledgments}
This work is supported by National Science Foundation (NSF), United States, Award Number: 2501703 and CRG grant from South Dakota Board of Regents (SDBOR).

%%%  REFERENCES  %%%%%%%%%%%%%%%%%%%%%%%%%%%%%%%%
%%
%% Put your references into your .bib file in the usual way. Run latex once, bibtex once, then latex twice.
%% The asmeconf.bst style allows @inproceedings and @proceedings to include: 
%%		venue = {Location of Conference}, 
%%		eventdate = {Month, days},

\nocite{*}%% <=== Delete this line unless you want to typeset the entire contents of your .bib file !!

\bibliographystyle{asmeconf}  %% .bst file following ASME conference format. Do not change.
\bibliography{references}%% <=== change this to the name of your bib file

\end{document}